# Big Bang Cosmology and Religious Thought

Jean-Pierre Luminet
Aix-Marseille Université, CNRS, LAM, Marseille, France

## Abstract

Alexander Friedmann and Georges Lemaître are undoubtedly the real fathers of Big Bang cosmologies. In this article I study the influences their work underwent due to some religious as well as anti-religious ideas. During his career Lemaître faced criticisms coming from non-believing scientists, who reproached him (wrongly) to have developed his primeval atom cosmology for conciliatory reasons. In the former case of Friedmann, we know that his 1922 proposal of a "creation of the world out of nothing" was criticized by Einstein for metaphysical reasons. The essence of such hostile reactions is the fact that presumably "good" scientific contents should not be influenced by religious ideas. Nevertheless, metaphysical and theological ideas can play an important role inside the science research processes, as an epistemological tool helping to clarify the use of some fundamental notions.

## Introduction

The discovery of dynamical models of relativistic cosmology and the recognition of an evolution of the universe as a whole from a possible singular origin, now called "Big Bang", have their source in the theory of general relativity, whose field equations were given by Albert Einstein in 1915. These fundamental ideas are the work of two pioneers who, armed only with their "pen" and an intuition that can be described without emphasis as brilliant, unveiled this new vision of the world: the Russian Alexander Friedmann (1888-1925) and the Belgian Georges Lemaître (1894-1966). The image that cosmology offers today of the evolution of the universe is remarkably close, in its fundamental concepts, to the scheme initially proposed by Friedmann, and especially by Lemaître who, as early as 1931, also predicted that a correct description of the early universe should also rest on the other great pillar of modern physics: quantum mechanics.

Let us briefly summarize their groundbreaking contributions to modern cosmology. While Einstein created the theory of general relativity and wrote the field equations governing the physical and geometrical properties of the universe, in 1922 Friedmann discovered the non-static solutions of these equations, describing the time variation of space, and glimpsed its possible beginning into a singularity[1]; in 1927 Lemaître was the first one to link the theoretical concept of the expansion of space to the apparent motion of galaxies observed by Edwin Hubble and his collaborators[2], and in 1931 he laid the physical foundations of the Big Bang models, anticipating the fundamental role played by quantum mechanics and vacuum energy[3]; the same year Lemaître predicted a phase of accelerated expansion of the universe due to a kind of repulsive field of energy called the cosmological constant[4]. On this regard Friedmann and Lemaître can really be considered as the two fathers of Big Bang cosmology.

The purpose of the present essay is to study the influences their work underwent due to some religious as well as anti-religious ideas. As emphasized by the best specialist of the subject Dominique Lambert[5], during his career Lemaître faced criticisms coming from non-believing scientists, who reproached him (wrongly) to have developed his primeval atom cosmology for conciliatory reasons. In the former case of Friedmann, we know that his 1922 proposal of a "creation of the world out of nothing" was criticized by Einstein for metaphysical reasons.

The essence of such hostile reactions is the fact that presumably "good" scientific contents should not be influenced by religious ideas. Nevertheless, metaphysical and theological ideas can play an important role inside the science research processes, as an epistemological tool helping to clarify the use of some fundamental notions. For instance, Einstein refused the concept of creation because of his Spinozist philosophy, which excludes any form of creation in a theological sense, then this prejudice lead him to initially favor models without natural beginning.

---

[1] A.A.Friedmann, "Über die Krümmung des Raumes", Zeitschrift für Physik, vol. 10, 1922, p. 377-386. English transl. « On the curvature of space » by B. Doyle, in K.R. Lang et 0. Gingerich (éd.), *A Source Book in Astronomy and Astrophysics, 1900-1975,* Cambridge, Mass., Harvard University Press, 1979, p. 838-843.

[2] G. Lemaître, *Un univers homogène de masse constante et de rayon croissant, rendant compte de la vitesse radiale des nébuleuses extra-galactiques*, in Annales de la Société Scientifique de Bruxelles, série A, t. XLVII, avril 1927, pp. 29-39 (49-59). Engl. Transl. (by G. Lemaître) : *A homogeneous universe of constant mass and increasing radius accounting for the radial velocity of extra -galactic nebulae*, in Monthly Notices of the Royal Astronomical Society, t. XCI, 1931, pp. 483-490.

[3] G. Lemaître, *The beginning of the world from the point of view of quantum theory,* in Nature, t. CXXVII, 1931, p. 706.

[4] G. Lemaître, *The expanding universe*, in Monthly Notices of the Royal Astronomical Society, t. xci, mars 1931, pp. 490-501.

[5] D. Lambert, "Religious interferences at the origin of cosmology? The case of Georges Lemaître", *Forum* vol.4 (2018) 149-168. The relations between science and religious faith in Lemaître's thought have also been analyzed in O. Godart, "Contributions of Lemaître to General Relativity (1922-1934)", op. cit. §7; O. Godart and M. Heller, "Les relations entre la science et la foi chez Georges Lemaître", in *Commentarii, Pontifical Academy of Sciences,* vol. III, n°121, p.1-12; D. Lambert, " Mgr. Georges Lemaître et les "Amis de Jésus" ", *Revue Théologique de Louvain,* vol. 27, 1996, 309-343, and " Monseigneur Georges Lemaître et le débat entre la cosmologie et la foi ", op. cit. Lemaître himself expressed on the question in "La culture catholique et les sciences positives", *Actes du VIe congrès catholique de Malines,* t. V, pp. 65-70.

**The creation of the world according to A. Friedmann**

As far as the cosmic singularity (i.e. the universe reduced to a point) is concerned, Friedmann posed for the first time the problem of the beginning and the end of the universe in scientific terms[6]. He could not help but see a metaphysical implication when he wrote[7]:

> One can remember here Hindu mythology about life periods; there is also the possibility of *"creation of the world from nothing"*. But one must look now on all this as on some curious facts which can't be confirmed by nonsufficient astronomical data ".

The concepts of cyclic or oscillatory universes are indeed quite frequent in mythology. Among Hindus, each cycle of the universe is a *kalpa*, or day of Brahmâ, which lasts 4320 million years. Vishnu, who controls the universe, has a life of one hundred "cosmic years", each containing 360 days of Brahmâ. After 36,000 cycles, corresponding to about 150 trillion earth years, the world comes to an end, and only the Spirit survives. After an indeterminate period of time, a new world and a new Vishnu emerge, and the cycle begins again. Note that the day of Brahmâ is of the same order as the age of the universe in modern cosmology.

On the other hand, the term "creation of the world out of nothing" he used[8], once launched in the field of relativistic cosmology, will cause many upheavals and misunderstandings, and will psychologically block most physicists. This problem, which Friedmann judged as a "curiosity" - perhaps out of modesty, or to distance oneself once again from hazardous theological interpretations - has, however, become the major challenge of contemporary research, as it is closely linked to the unsolved problem of quantum gravity[9].

In the general bibliography of A. Friedmann, we note the existence of a lost manuscript, precisely entitled *Creation* (*Mirozdanie*). No one knows what its contents might have been, but it is not impossible that Friedmann developed a theological point of view in it - a point of view that he refused to address in his official publications, as he pointed out on several occasions. Talking about the creation of the universe under the communist regime was somewhat daring politically, although Friedmann did not care much about politics. It was only in the 1960s that Soviet science became converted to the Big Bang concept[10]. There are indications that if the Russian scientist had survived his theory longer, he might have been imprisoned and persecuted[11].

---

[6] On the modern point of view concerning cosmic singularity - for example, whether it is an inevitable consequence of general relativity, whether it took place at a single point or several, etc. - the reader may consult S.W. Hawking and G.F.R. Ellis, *The Large Scale Structure of Spacetime* (see Bibliography), and, at a less technical level, S. Hawking and R. Penrose, *The Nature of Space and Time,* Princeton University Press, 1996.

[7] *The world as space and time*, English translation by V. Petkov, Minkowski Institute Press (2016).

[8] A reference to the *Second Book of the Maccabees*, **7**, 28.

[9] Some conjectures on possible solutions can be found in L. Smolin, *Three Roads to Quantum Gravity,* Weidenfeld & Nicolson, 2014, and J.-P. Luminet, *L'écume de l'espace-temps*, Odile Jacob, 2020.

What do we know about religious beliefs of Friedmann?[12]. He was "officially" an Orthodox Christian, but there are some arguments that suggest a nonconformal attitude to his Christianity. The most early in time argument is that the only discipline at the St. Peterburg gymnasium on which the young pupil Alexander Friedmann had mainly excellent marks (5/5) was catechism, whereas at the same time his marks in mathematics were only 3 (i.e., "satisfactory"). Friedmann married his first wife E.N. Dorofeeva in the church, but his colleague V.V. Doynikova remembered many of his words about his religious mindedness. Some of them were of mystical kind. Friedmann "was found of the 'occult' in general though that he can cure toothache by words", etc.

Deep religious were people closed to Friedmann, for example Vladimir Ivanovich Smirnov, the famous Russian mathematician. On the grave of Friedmann in the Smolenskoie graveyard of St. Petersburg there is the Orthodox cross. May be this was done according to his last will, or made by his relatives who knew about his attitude to faith. In 1925, just before his death, being the Director of the Main Geophysical Observatory in the time of antireligious and antichurch persecutions of Communist power, he married according to Orthodox Christian rule his second wife N.E. Malinina in the church of Simferopol (Crimea). So the phrase "creation of the world" was not only a word for him.

One can say that in his work Friedmann realized the idea of F.M. Dostoevsky in *The Brothers Karamazov*, that only a non-Euclidean mind might be able to solve the contradiction between science and the religious revelation. V.V. Doynikova remembered that Dostoevsky was one of the favorite writers for Friedmann. One can remember also the well-known words of Einstein: "Dostoevsky gave me more than Gauss"!

**Georges Lemaître, priest and scientist**

---

[10] So much so that one of its most fervent supporters, Yacov B. Zeldovich, declared in 1982, before the International Astronomical Union, that the Big Bang was "as certain as the fact that the Earth revolves around the Sun". He ignored the ironic remark of his compatriot Lev Landau that "cosmologists are often wrong, but never in doubt"!

[11] Andrey Grib, private communication.

[12] Andrey Grib, *Early expanding universe and elementary particles (Appendix)*, Friedmann Laboratory Pub. Ltd., 1995.

The second "father" of Big Bang theory, Georges Lemaître, was a great scientisit but he was also a catholic priest with a deep fath. Thus it is legitimate to wonder about the possible interferences between his scientific practice and his religious commitment.

Georges Lemaître was born on July 17, 1894 in Charleroi, Belgium[13]. The eldest of a middle-class catholic family, he was educated at the Jesuit College in his native town, and rapidly felt a call to become both a priest and a scientist. However, his father advised him to complete engineering studies first before entering the seminary. The young Lemaître followed his father's recommendation and at the age of 17 he enrolled at the University of Louvain, where he studied engineering for three years and got his Bachelor degree. Just after, the First World War burst out. He voluntarily joined the Belgian Infantry and mainly Artillery, participating in important battles along the river Yser. It's at this time that Georges Lemaître chose his dual scientific and religious vocation. "There were two ways to get to the truth, and I decided to take them both," he would later tell an American journalist[14], but in this young age Lemaître was briefly tempted to unify science and religion. Lemaître dedicated time for prayer and for reading science books, such as *Electricité et Optique* of Henri Poincaré. In parallel he meditated the *Book of Genesis*, the *Psalms* and read carefully some books of Léon Bloy (1846-1917), a famous catholic writer who was a close friend of the Christian poet Charles Péguy and of the neo-Thomistic philosopher Jacques Maritain. Lemaître was then tempted to build a unified religious and scientific synthesis. In the trenches he wrote a short essay entitled *Les trois premières paroles de Dieu* (*The three first words of God*), an attempt to give a kind of exegesis of the *Book of Genesis*'s three first verses in a concordist-like style. He sent this essay to Léon Bloy and in 1918 he met him, in Bourg-la-Reine, near Paris, during his furlough. Bloy, who died shortly afterwards, gave him the advice to avoid any form of concordism, namely the attempt of mixing on the same level scientific and theological contents. Lemaître was a little bit puzzled, but he followed this profound advice and afterwards, during all the rest of his sacerdotal and academic life, he will adopt a position making a careful distinction between Science and Faith, as we shall see below in more details.

Returning to university in 1919, Lemaître moved from engineering studies to the much more abstract physical and mathematical sciences under the supervision of the famous mathematician Charles de la Vallée Poussin. There he proved to be a major force. In 1920, he obtained his doctorate in mathematics (today corresponding to the American Master's degree), while pursuing his vocation to the priesthood.

---

[13] For this biographical summary, I closely follow the excellent summary of Odon Godart, "Monseigneur Lemaître, sa vie, son œuvre", *Revue des Questions Scientifiques,* vol. 155, 1984, p. 155-182. For a more complete biography, D. Lambert, *Un atome d'univers. La vie et l'œuvre de Georges Lemaître.* Bruxelles, Lessius (2000). Engl. transl. L. Ampleman : *The Atom of the Universe. The Life and Work of Georges Lemaître*, Kracow, Copernicus Center Press 2015.

[14] Interview with the *New York Times Magazine*, February 19, 1933.

After his ordination in 1923, he belonged to a sacerdotal fraternity founded by Cardinal Mercier called "Les Amis de Jésus " (The Friends of Jesus), in which priests pronounced some vows unusual for secular priests, for example the vow of poverty. All along his life he will keep good relations with his hierarchy. In 1935, he will be made honorary Canon of Saint-Rombaut Cathedral in Mechelen (Malines) and one year later, Pope Pius XI will choose him in the first list of members of the Pontifical Academy of Sciences. In 1960 he will be appointed by Pope John XXIII as President of the Pontifical Academy of Sciences, and on that occasion elevated to the rank of Domestic Prelate.

### From Space Expansion to the Beginning of the World

But let's go back to his major scientfic achievements. In 1927, Georges Lemaître was the first physicist who gave the explanation of what we call now the Hubble Law saying that the speeds of the distant galaxies are proportional to their distances, up to a constant called now the Hubble constant. He explained that using a cosmological model, solution of Einstein equations of general relativity, representing an expanding universe. Lemaître had become familiar to the solutions of Einstein equations in 1920-23 but above all during his stay in Cambridge, UK (1923-1924) and at the MIT (1924-1925). He got the intuition of an expanding universe while working on the empty universe model of de Sitter. He had gotten the opportunity to collects some important and up to date data concerning the speeds and the distances of the galaxies during a travel through the USA, where he visited the main observatories of his time. Using a particular universe model (corresponding to a spherical and massive universe undergoing an exponential expansion without beginning nor end), he explained the redshifts of the galaxies (which measure their speeds) not as a real propermotion but as the consequence of space expansion. He also derived the expression and a provisional value of the co-called "Hubble constant". No religious or theological ideas motivated this explanation, which proceeded here of a logical dynamics inner to physics and astronomy, without out-of-the-field interference.

Lemaître published his paper on expanding space in 1927. Immediately he had to suffer objections. The strongest one went from Einstein himself. He had read the paper which was given to him by a friend. From 24 to 29 October 1927 the Fifth Solvay Conference in Physics took place in Brussels, one of the great meetings of world science. The Solvay Conference was devoted to the new discipline of quantum mechanics, whose problems disturbed many physicists. Among them was Einstein. For Lemaître, it was the opportunity to discuss with the father of general relativity. He later reported himself on this meeting:

> While walking in the alleys of the Parc Léopold, [Einstein] spoke to me about an article, little noticed, which I had written the previous year on the expansion of the universe and which a friend had made him read. After some favorable technical remarks, he concluded by saying that from the physical point of view that appeared completely abominable to him. As I sought to prolong the conversation, Auguste Piccard, who accompanied him, invited me to go up by taxi with Einstein, who was to visit his laboratory at the University of Brussels. In the taxi, I spoke speeds of nebulae and I had the impression that Einstein was hardly aware of the astronomical facts. At the university, all occurred in German[15].

In his tasty notes added to Odon Godart's article[16], André Deprit (a former student of Lemaître) gives a more picturesque and slightly different version of this encounter. In particular, he states that Lemaître did not know German, which may explain why the Belgian scholar did not cite Friedmann's earlier work in his 1927 article.

Einstein's response to Lemaître shows the same unwillingness to change his position that characterized his former response to Friedmann: he accepted the mathematics, but not a physically expanding universe. According to D. Lambert[17], this reaction came from the fact that the Einstein's implicit philosophy was inspired by Spinoza. For the Dutch philosopher, "God" (Deus) was identified with the "Nature" (Natura): "*Deus sive Natura*". Consequently, due to the immutability of God, one could not accept any motion or evolution of the Nature itself (*Ethics* II, scolie of Lemma VII). Einstein thus rejected the idea of an evolving universe, i.e. a world with a real history. This "theological" prejudice lead him also to criticize strongly the idea of expanding (and contracting) universes put forward by Friedmann and Lemaître.

The year 1931 can undoubtedly be called the Georges Lemaître's *annus mirabilis*. Indeed, major contributions to relativistic cosmology by the Belgian physicist and priest appeared within a few months :

---

a) the English translation of his 1927 article in the M.N.R.A.S., sponsored by Arthir Eddington who wanted to introduce the Lemaître's ideas on dynamical universes to a large communauty

b) *The expanding universe*[18], in which Lemaître calculated that the expansion of space could be induced by a preceding phase of « stagnation » taking place about $10^{10}$ years in the past, itself preceded by a singular origin called the primeval atom, that occured a finite time in the past

c) the short note *The beginning of the world from the point of view of quantum theory*, published in the March 21 issue of *Nature,* that can be considered as the true "chart" of the Big Bang theory[19] (ref).

Since the creation of the universe a finite time ago is a dogma in Christian thought, it might be tempting to jump to the conclusion that this time, the Lemaître's model of an explosive universe was motivated by the aim to reconcile relativistic cosmology with religious belief. But again it was not the case. The original manuscript (typed) version of the 1931 Lemaître's article in *Nature*, preserved in the Archives Lemaître at the University of Louvain, ended with a sentence crossed out by Lemaître himself and which, therefore, was never published. Lemaître initially intended to conclude his letter to *Nature* by

> I think that every one who believes in a supreme being supporting every being and every acting, believes also that God is essentially hidden and may be glad to see how present physics provides a veil hiding the creation.

This well reflected his theological view of a hidden God, not to be found as the Creator at the beginning of the universe. But before sending his paper to *Nature*, Lemaître probably realized that such a reference to God could mislead the readers and make them think that his hypothesis gave support to the Christian notion of God.

As well analyzed by the best specialist of the subject Dominique Lambert[20], Lemaître will preserve all his life the conception of a supreme and inaccessible

---

[18] Not to be confounded with *L'Univers en expansion*, Annales de la Société Scientifique de Bruxelles A 53, 51 (1933), reproduced as a Golden Oldie as *The expanding Universe,* (translated by M. A. H. MacCallum), Gen. Relativ. Gravit. 29, n∘5, 641 (1997). Editorial note by A. Krasinski, p. 637.

[19] For a detailed analysis see J.-P. Luminet, "Editorial note to 'The beginning of the world from the point of view of quantum theory'", *Gen. Rel. Grav.* Vol 43(10), pp. 2911-2928 (2011).

[20] D. Lambert, "Religious interferences at the origin of cosmology? The case of Georges Lemaître"*, Forum* vol.4 (2018) 149-168. The relations between science and religious faith in Lemaître's thought have also been analyzed in O. Godart, "Contributions of Lemaître to General Relativity (1922-1934)", op. cit. §7; O. Godart and M. Heller, "Les relations entre la science et la foi chez Georges Lemaître", in *Commentarii, Pontifical Academy of Sciences,* vol. III, n°121, p.1-12; D. Lambert, " Mgr. Georges Lemaître et les "Amis de Jésus" ", *Revue Théologique de Louvain,* vol. 27, 1996, 309-343, and " Monseigneur Georges Lemaître et le débat entre la

God of which the prophet Isaiah speaks, enabling him to keep the natural origin of the world within the strict limits of physics, without mixing it with a supernatural creation. As a priest just like a scholar in theology, Lemaître was very conscious of the potential conflict – or, on the contrary, of the concordance – between the Christian dogma of a world created by God and the scientific theory of a universe formed approximately ten billion years ago.

Contrary to some other Christian cosmologists[21], Lemaître took guard not to use one of these two "ways of knowledge" as a legitimation of the other. He had for example great care to distinguish between the "beginning" and the "creation" of the world, and never spoke about the initial state of the universe in terms of "creation" (contrary to Friedmann, who eventually appears more « concordist » than the Belgian priest). Lemaître was convinced that science and theology dealt with two separated worlds, and that the scientific cosmology of the Big Bang neither confirmed nor refuted the Christian notion of God's creation of the world.

### Misunderstandings about the Primeval Atom Hypothesis

Lemaître's belief in two separate levels of understanding, one scientific and the other religious, did not imply, however, that he found cosmology irrelevant to religious thought. He believed that, on a broader ethical level, religious and philosophical values are important, even essential, to the scientist, but that they should not interfere with his methods or conclusions. With the humor he often displayed, he wrote in 1934 that searching for truth implies a search for the soul as well as for "spectres" (those of extragalactic nebulae)!

In an interview given in 1933 upon his return to the United States and reproduced in a long article in the New York Times[22], Lemaître clearly stated his conception of the relationship between science and religion. We reproduce here the most significant extracts:

> There is non conflict between science and religion […] Here we have this wonderful, this incessantly interesting and exciting universe. When we try to learn more about it, learn how it began and how it is put together, to find what it is all about, as you say in America, what are we doing? Only seeking the truth. And is not truth-seeking a service to God? Certainly everything in the Bible and in all authoritative Christian doctrine teaches that it is. Has any logical religious thinker of any faith ever denied it?

[...]

Once you realize that the Bible does not purport to be a textbook of science, the old controversy between religion and science vanishes. There is no reason to abandon the Bible because we now believe that it took perhaps ten thousand million years to create what we think is the universe. Genesis is simply trying to teach us that one day in seven should be devoted to rest, worship and reverence—all necessary to salvation.

As a matter of fact, neither St. Paul nor Moses had the slightest idea of relativity. The writers of the Bible were illuminated more or less—some more than others—on the question of salvation. On other questions they were as wise or as ignorant as their generation. Hence it is utterly unimportant that errors of historic and scientific fact should be found in the Bible, especially if errors relate to events that were not directly observed by those who wrote about them. The idea that because they were right in their doctrine of immortality and salvation they must also be right on all other subjects is simply the fallacy of people who have an incomplete understanding of why the Bible was given to us at all.

[...]

The real cause of conflict between science and religion is to be found in men and not in the Bible or the findings of physicists. When men were told that they had the right to interpret the Bible's teachings according to their own lights, naturally some were bound to decide that its science was infallible and others that it did not agree with modern instrumental measurements and was proof of opposite doctrines. The conflict has always been between those who fail to understand the true scope of either science or religion. For those who understand both, the conflict is simply about descriptions of what goes on in other people's minds.

Lemaître will also express this more "officially" in his speech at the 1958 Solvay congress, where he mentioned that the Big Bang theory remained totally outside of any metaphysical or religious question.

In the meantime an important episode happened, namely the 1951 famous address *Un' Ora* (The proofs of the Existence of God in the Light of Modern Natural Sciences), given by Pope Pius XII in front of the Pontifical Academy of Sciences, which implicity referred to the primeval atom hypothesis but without quoting explicitly Lemaître:

Contemporary science, with one sweep back across the centuries, has succeeded in bearing witness to the august instant of the primordial *Fiat Lux,* which along with the matter there burst forth from nothing a sea of light and radiation . . . Thus, with that concreteness which is characteristic of physical proofs, modern science has confirmed the contingency of the universe and also the well-founded deduction to the epoch when the world came forth from the hands of the creator.

As well explained by Dominique Lambert[23], the pontifical talk is often read as a defense of the primeval atom hypothesis with a concordist-like style. In fact, the address wanted to show that the proofs of God's existence of St Thomas Aquinas can be revitalized by founding new supports in contemporary science. The *Un'Ora* address suggests that physics (thermodynamics, nuclear physics, cosmology) can bring some data to defend the mutability of the universe and then to give a new foundation to initiate the Thomistic proof of the existence of God based on motion, i.e. change. This was explicitly noted by the thomistic philosopher of Louvain Fernand Van Steenberg - a colleague of Lemaître, together with him member of "Les Amis de Jésus". In his book entitled "*Dieu caché. Comment savons-nous que Dieu existe?*" (*Hidden God. How do we know that God exists ?*), he wrote, referring to the address *Un'Ora* :

> Pius XII was deeply struck and obviously delighted by the recent discoveries of physics and by the new orientation they give to the cosmogonic theories of scientists. Far from contradicting the theses of traditional philosophy or the data of Christian revelation, physics reveals facts which reinforce the empirical starting points of the philosophical proofs of the existence of God [. . .] However, as one may have noticed, each time his admiration for the discoveries of science risks leading him to imprudent declarations, Pius XII stops in time and brings the necessary nuances and reservations: on their own, he says, science cannot prove the existence of God; it is when he thinks as a philosopher that the scientist succeeds[24].

Georges Lemaître thought exactly the same thing, and he was afraid that the Pope's address could suggest to many astronomers that his cosmological hypothesis of the primeval atom had been conceived for apologetic reasons. Furthermore, he knew better than anyone that the hypothesis was not yet confirmed by observations. In 1952, stopping at the Vatican on a trip to South Africa, he succeeded in advising the Pope, who had to deliver a speech to the Eight General Assembly of the International Astronomical Union, not to refer to the primeval atom. For this Lemaître consulted with two men, Father O'Connell, a science advisor to Pius XII, and the Cardinal Secretary of State. Lemaître's short visit had the intended effect. The new Pope's speech primarily praised the advances in astrophysics research in the last fifty years, making

only a brief statement on the Big Bang, namely that

> The human spirit, upon considering the vast paths traveled by galaxies, becomes in some manner a spectator at the cosmic events that occurred on the very morning of creation[25].

And Pius XII never mentioned the primeval atom hypothesis again. In a conference in 1963 entitled "Univers et Atome", Lemaître declared, concerning the papal address:

> It is clear that the attitude of the Pontiff is on his own ground and has no relation to Eddington's or my theories. Moreover, my name is not mentioned in the Pope's speech.

For Lemaître, creation was not a question of beginning. In a text edited long after his death but written in the end of the thirties, he said[26]:

> What happened before that? Before that we have to face the zero value of the radius (of the universe). We have discussed how far it had to be taken as strictly zero, and we have seen that it means a very trifling quantity, let us say few light-hours. We may speak of this as of a beginning; I do not say a creation. Physically it is a beginning in the sense that if something had happened before it, it has no observable influence on the behavior of our universe, as any feature of matter before this beginning has been completely lost by the extreme contraction at the theoretical zero. A pre-existence of the universe has a metaphysical character. Physically, everything happens as if the theoretical zero was really a beginning. The question if it was really a beginning or rather a creation: something starting from nothing, is a philosophical question that cannot be settled by physical or astronomical considerations.

In his beautiful memorial article on Lemaître written for the Pontifical Academy of Sciences in 1968, Paul Dirac recalls that one day they were chatting about cosmic evolution and that, feeling stimulated by the grandeur of the image the Belgian scholar-priest had given, he told him that he thought cosmology was the branch of science that was closest to religion. But Lemaître did not agree with Dirac. After careful consideration, he suggested that the scientific discipline closest to religion was rather psychology.

---

[25] Elio Gentili and Ivan Tagliaferri, *Science and Faith — The Protagonist Priest and Religious Scientist* (Rome, Instituto Geografico de Agostini), 287.
[26] G. Lemaître, *The expanding universe: Lemaître's unknown manuscript* (introduction by O. Godart and M. Heller). Tucson (Arizona): Pachart Publishing House (1985), p.47.